\documentstyle[sprocl]{article}

\input{epsf.sty}

\def \tablerule{\noalign {\vskip3truept\hrule\vskip3truept}}

\def\gsim{\mathrel{\rlap{\raise 0.511ex \hbox{$>$}}{\lower 0.511ex\hbox{$\sim$}}}}
\def\lsim{\mathrel{\rlap{\raise 0.511ex \hbox{$<$}}{\lower 0.511ex\hbox{$\sim$}}}}

\def\ra{\rightarrow}

\def\be{\begin{equation}}
\def\ee{\end{equation}}
\def\bea{\begin{eqnarray}}
\def\eea{\end{eqnarray}}

\begin{document}

\title{MODELLING GRAVITATIONAL WAVES FROM INSPIRALLING COMPACT BINARIES}

\author{T. Damour$^1$, B.R. Iyer$^2$ \underline{B.S. Sathyaprakash$^3$}}
\address{$^1$ Institut des Hautes Etudes Scientifiques, 91440
Bures-sur-Yvette, France \\
$^2$ Raman Research Institute, Bangalore 560 080, India\\
$^3$ Cardiff University of Wales, P.O. Box 913, Cardiff, CF2 3YB, U.K.}

\maketitle
\begin{abstract}
Gravitational waves from inspiralling compact binaries can
be reliably extracted from a noisy detector output only
if the template used in the detection is a faithful representation
of the true signal. In this article we suggest a new approach to
constructing faithful signal models. 
\end{abstract}

\section {Introduction}

In searching for gravitational waves from an inspiralling compact binary 
(ICB) we are
faced with the following data analysis problem: On the one hand, we have
some exact gravitational wave form $h^X (t;\lambda_k)$ where $\lambda_k$,
$k=1, \ldots, n_{\lambda},$ are the parameters of the signal (eg.,
the masses $m_1$ and $m_2$ of the members of the emitting binary).  
On the other hand, we have theoretical calculations of the motion 
of~\cite{DD81}, and
gravitational radiation from~\cite{BDIWW,BDI,WW,BIWW,B96}, 
binary systems of compact bodies (neutron stars or black holes) giving
the post-Newtonian (PN) expansions (expansions in powers of $v/c$) of
an energy function $E(v)$ and a gravitational 
wave luminosity function $F(v)$. Here, the dimensionless
argument $v$ is an invariantly defined ``velocity''
related to the instantaneous {\it gravitational wave}
frequency $f^{\rm GW}$ ($=$ twice the {\it orbital} frequency) by
$ v = (\pi \, m \, f^{\rm GW})^{\frac{1}{3}},$
where $m \equiv m_1 + m_2$ is the total mass of the binary. 
Given the energy and flux functions one needs to compute
the ``phasing formula'', i.e. an accurate mathematical
model for the evolution of the gravitational 
wave phase $\phi^{\rm GW} = 2\Phi = F[t;\lambda_i],$
involving the set of parameters $\{ \lambda_i \}$ carrying 
information about the emitting binary system.
The standard energy-balance equation $dE_{\rm tot} / dt =-F$ gives the 
following parametric representation of the phasing formula:
\begin{equation}
t(v) = t_c + m \int_v^{v_{\rm lso}} dv \, 
\frac{E'(v)}{F(v)},\ \  
\Phi (v) = \Phi_c + \int_v^{v_{\rm lso}} dv' v'^3 \, 
\frac{E'(v')}{F(v')},
\label {eq:6}
\end{equation}
where $t_c$ and $\Phi_c$ are integration constants. 
We now turn to the discussion of what is known about 
the two functions $E(v)$ and $F(v)$ entering the phasing 
formula and how that knowledge can be improved.

\section {New Energy and Flux Functions}

Let $E_{T_n}\equiv \sum_{k=0}^n E_k(\eta)v^n$ and 
$F_{T_n}\equiv \sum_{k=0}^n F_k(\eta)v^n,$ where 
$\eta\equiv m_1 m_2/m^2$ is the symmetric mass ratio, 
denote the $n^{\rm th}$-order 
Taylor approximants of the energy and flux functions.
For finite $\eta,$ the above Taylor approximants are 
known~\cite{BDIWW,BIWW,B96} for $n\le 5.$
In the test mass limit, i.e. $\eta \ra 0,$  $E(v)$
is known exactly, the exact flux is known numerically~\cite{P95,TTS97}
and analytically the flux 
 is  known~\cite{TTS97,CFPS,P93} up to the order $n=11$.
The problem is to construct a sequence of approximate wave forms $h_n^A
(t;\lambda_k)$, starting from the PN expansions of $E(v)$
and $F(v).$ In formal terms, any such construction defines a {\it map} from
the set of the Taylor coefficients of $E$ and $F$ into the (functional) space
of wave forms. Up to now, the literature has only considered the
standard map, say $T$,
\begin{equation}
(E_{T_n} ,F_{T_n}) \stackrel{T}{\rightarrow} h_n^T (t,\lambda_k) \, , \label
{eq:n4}
\end{equation}
obtained by inserting the successive Taylor approximants
into the phasing formula~\cite{Cutler,P95}.  
We propose a new map, say ``$P$'', based on two essential ingredients:
(i) the introduction, on theoretical grounds, of two new, supposedly 
more basic and hopefully better behaved, energy-type and flux-type 
functions, say $e(v)$ and $f(v)$, and (ii) the systematic use of 
Pad\'e approximants (instead of straightforward Taylor
expansions) when constructing successive approximants of the intermediate
functions $e(v)$, $f(v)$. Schematically, our procedure is~\cite{dis97}:
\begin{equation}
(E_{T_n} , F_{T_n}) \rightarrow (e_{T_n} , f_{T_n}) \rightarrow (e_{P_n} ,
f_{P_n}) \rightarrow (E[e_{P_n}] , F[e_{P_n} , f_{P_n}]) \rightarrow h_n^P
(t,\lambda_k) \, . \label {eq:n5}
\end{equation}

Our new energy function $e(x),$ where $x\equiv v^2,$ 
is constructed out of the total relativistic
energy $E_{\rm tot}$ using
\begin{equation}
e(x) \equiv \left( \frac{E_{\rm tot}^2 - m_1^2
-m_2^2}{2m_1 m_2}\right)^2 -1 \, . \label{eq:N9}
\end{equation}
The function $e(x)$ is symmetric in the two masses. 
The function $E(x)$ entering the phasing formulas is given in terms
of $e(x)$ by
\begin {equation}
E(x) = \left [1 + 2 \eta
\left (\sqrt {1+e(x)} - 1\right ) \right ]^{1/2} - 1.
\label{eq:Etilde1}
\end{equation}

In the test-mass limit the exact expression for the function $e(x)$ can
be computed from its definition above which when substitute in
Eq.~(\ref{eq:Etilde1}) gives the well known energy function for 
a test mass in orbit around a Schwarzschild black hole:
\begin{equation}
e(x)=-x \frac{1-4x}{1-3x},\ \ 
E'(x)=-\eta \sqrt{x} \frac {1-6 x}{(1-3x)^{3/2}}. 
\end{equation}
The test mass exact energy function $e(x)$ has a simple pole 
singularity while the function $E(x)$ has in addition a branch 
cut. Therefore the function $e(x)$ is more suitable in analysing 
the analytic structure.
In the comparable mass case, on the grounds of mathematical
continuity between the case $\eta \rightarrow 0$ and the case 
of finite $\eta,$ one can expect the exact
function $e(x)$ to admit a simple pole singularity on the real axis 
$\propto (x-x_{\rm pole})^{-1}.$ We do not know the
location of this singularity, but {\it Pad\'e approximants} are
excellent tools for giving accurate representations of functions having
such pole singularities~\cite{BO}. 
Indeed, it turns out that the Pad\'e approximant
of the 2PN expansion of $e(x)$ gives the exact energy
function~\cite{dis97}.  We therefore use Pad\'e approximants of
$e(x)$ in computing the energy function $E(x)$ entering the phasing formula
instead of the standard Taylor approximants. This greatly improves the
accuracy of the phasing formula.

It has been pointed out~\cite{CFPS} that the flux function 
$F(v;\eta =0)$ has a simple pole at the light ring $v^2 = 1/3$. 
The light ring orbit corresponds to a simple pole $x_{\rm pole}
(\eta)$ in the
new energy function $e(x;\eta)$. Let us define the corresponding (invariant)
``velocity'' $v_{\rm pole} (\eta) \equiv \sqrt{x_{\rm pole} (\eta)}$. This
motivates the introduction of the following ``factored'' flux function,
its Pad\'e approximants $f_{P_n},$ and the corresponding flux function
entering the phasing formula:
\begin{equation}
f(v;\eta) \equiv \left( 1-v/v_{\rm pole} \right) \, F (v;\eta),\ \ 
\label{eq:n33}
F_{P_n}(v;\eta)\equiv \left(1-v/v_{\rm pole} \right )^{-1}\, f_{P_n} (v;\eta). 
\label{eq:nn36}
\end{equation}

\section {Effectual and Faithful Signal Models}

In order to test whether a given approximant to the wave form is good
or not we make use of the statistic used in detecting the ICB signal.
We shall say that a multi-parameter family of approximate wave
forms $h^A (t;\mu_k)$, $k=1,\ldots ,n_{\mu,}$ is an {\it effectual} model of
some exact wave form $h^X (t;\lambda_k)$; $k=1,\ldots ,n_{\lambda}$ (where one
allows the number of model parameters $n_{\mu}$ to be different from, i.e. in
practice, strictly smaller than $n_{\lambda}$) if the overlap, or normalized
ambiguity function, between $h^X (t;\lambda_k)$ and the time-translated family
$h^A (t-\tau ; \mu_k)$,
\begin{equation}
{\cal A} (\lambda_k ,\mu_k) = \max_{\tau,\phi} \frac{ \langle h^X
(t;\lambda),h^A (t-\tau ;\mu)\rangle}{\sqrt{\langle h^X (t;\lambda),h^X
(t;\lambda)\rangle \langle h^A (t;\mu),h^A (t;\mu)\rangle}} \, , \label {eq:n6}
\end{equation}
is, after maximization on the model parameters 
$\mu_k$, larger than some given threshold, 
e.g. $\max_{\mu_k} {\cal A} (\lambda_k , \mu_k) \geq
0.965$~\cite {eventrate}. 
[In Eq. (\ref{eq:n6}) the scalar product $\langle h,g \rangle$ denotes
the usual Wiener bilinear form involving the noise
spectrum $S_n (f)$.] While an {\it effectual} model may be a
precious tool for the successful detection of a signal, 
it may do a poor job in
estimating the values of the signal parameters $\lambda_k$. 
We shall then say
that a family of approximate wave forms $h^A (t; \lambda_k^A)$, where the
$\lambda_k^A$ are now supposed to be in correspondence with (at least a subset
of) the signal parameters, is a {\it faithful} model of $h^X (t;\lambda_k)$ if
the ambiguity function ${\cal A} (\lambda_k ,\lambda_k^A)$, Eq. (\ref{eq:n6}),
is maximized for values of the model parameters $\lambda_k^A$ which differ from
the exact ones $\lambda_k$ only by acceptably small biases.
A necessary criterion for faithfulness,
and one which is very easy to implement in practice, is that the ``diagonal''
ambiguity ${\cal A} (\lambda_k ,\lambda_k^A = \lambda_k)$ be larger than, say,
0.965. Using this terminology Eq. (\ref{eq:n5})
defines approximants which, for practically all values of
$n$ we could test, are both more effectual (larger overlaps) and more faithful
(smaller biases) than the standard approximants Eq. (\ref{eq:n4}).
The new sequence of $P$-approximants 
exhibit a systematically better convergence behavior 
than the $T$-approximants~\cite{dis97}.  The overlaps they achieve at a 
fixed PN order are usually much higher. From our 
extensive study~\cite{dis97} of the formal ``test-mass limit" 
$\eta \equiv m_1m_2/(m_1+m_2)^2 \rightarrow 0,$
it appears that the presently known $(v/c)^5$-accurate 
PN results allow one to construct approximants having 
overlaps larger than 96.5\%. Such overlaps are enough to guarantee 
that no more than 10\% of signals may remain undetected.

\begin {table}[h]
\caption {Fraction of events ${\cal F}$ accessible relative to 
the case when the true signal is known, percentage bias in
the estimation of total mass ${\cal B}_m=100(1-m^A/m^X)$ and percentage bias in
the estimation of the mass ratio ${\cal B}_\eta=100(1-\eta^A/\eta^X),$ using
wave forms $h^T_n$ and $h^P_n$, respectively. ($A$ is either $T$ or $P,$
$X$ is for exact, and $n$ is the order of the approximant)}
\label {table:eventrate}
\begin {tabular}{ccccccc}
\tablerule
$n$ & ${\cal F}^T$ & ${\cal F}^P$ & ${\cal B}^T_m$ & ${\cal B}^P_m$ 
& ${\cal B}^T_\eta$ & ${\cal B}^P_\eta$\\
\tablerule
& \multicolumn{6}{c}{Neutron star-black hole binaries}\\
\tablerule
4 & 0.928 & 0.998 &$-6.96$ & $-3.61$  & $ 12.2$ & $5.64$   \\
5 & 0.945 & 1.000 &$-97.2$ & $-1.11$  & $ 69.3$ & $1.83$   \\
6 & 0.967 & 0.998 &$ 1.00$ & $-0.157$ & $-1.56$ & $0.263$   \\
\tablerule
& \multicolumn{6}{c}{Black hole-black hole binaries}\\
\tablerule
4 & 0.920 & 0.991 &$ 1.40$ & $-1.524$ &  $ 0.282$ & $0.700$   \\
5 & 0.532 & 0.998 &$-23.6$ & $-0.205$ &  $23.1$   & $0.042$   \\
6 & 0.988 & 1.000 &$0.391$ & $ 0.019$ &  $ 1.44$  & $0.119$\\   
\tablerule
\end{tabular}
\end{table}

Our results are summarized in Table~\ref{table:eventrate}, for
two archetypal binaries invovling neutron stars (NS) and
black holes (BH), where we have tabulated the fraction of events which
the templates constructed out of $T$- and $P$-approximants
would detect relative to the total number of events that
would have been detectable if we have had access to the true signal.
We have also listed biases in the measurement of parameters.
We clearly notice the superiority of the $P$-approximants. 

Though we believe that the the new approximants $h_n^P$ are superior
over the standard ones $h_n^T$ and shows the practical
sufficiency of the presently known $v^5$-accurate
PN results, we still think that it is an important (and
challenging) task to improve the (finite mass) PN results. 
Our calculations
also suggest that knowing $E$ and $F$ to $v^6$ would further improve the
effectualness (maximized overlap larger than 99.5\%) and, more importantly, the
faithfulness (diagonal overlap larger than 98\%) to a level allowing a loss in
the number of detectable events smaller than 1\%, and significantly smaller
biases (smaller than 0.5\%) in the parameter estimations than 
the present $(v/c)^5$ results (about 1---5\%).

\end{document}